\documentclass[a4paper]{article}
\usepackage{amsmath,amstext,amsgen,amsbsy,amsopn,amsfonts,amssymb}
\usepackage{easybmat}
\usepackage{graphics}
\usepackage[pdftex]{graphicx}
\usepackage{epsfig}
\usepackage{amsthm}
\usepackage{bm}
\usepackage{algorithm}
\usepackage{algorithmic}

\usepackage{wrapfig}
\usepackage{hyperref}

\begin{document}
\large
\title{\textbf{Estimating the Number of Infected Cases in COVID-19 Pandemic}}
\author{
Donghui Yan$^{\dag}$, 
Ying Xu$^{\ddag}$, Pei Wang$^{\P}$
\vspace{0.1in}\\
$^\dag$Department of Mathematics and Program in Data Science\\
University of Massachusetts, Dartmouth, MA 02747\\[0.04in]
$^\ddag$Indigo Agriculture Inc, Boston, MA\\[0.04in]
$^{\P}$Icahn School of Medicine at Mount Sinai, NYC, NY
}

\date{\today}
\maketitle \normalsize

\begin{abstract}
\noindent
The COVID-19 pandemic has caused major disturbance to human life. An important reason behind the widespread 
social anxiety is the huge uncertainty about the pandemic. A fundamental uncertainty is how many or what percentage 
of people have been infected. There are published and frequently updated data on various statistics of the pandemic, 
at local, country or global level. However, due to various reasons, many cases were not included in those reported numbers. We 
propose a structured approach for the estimation of the number of unreported cases, where we distinguish cases that arrive late 
in the reported numbers and those who had mild or no symptoms and thus were not captured by any medical system at all. We use 
post-report data for the estimation of the former and population matching to the latter. We estimate that the reported number of infected cases 
in the US should be corrected by multiplying a factor of 220.54\% as of Apr 20, 2020, while the infection ratio out 
of the US population is estimated to be 0.53\%, implying a case mortality rate at 2.85\% which is close to the 3.4\% 
suggested by the WHO in Mar 2020. Towards the end of the summer of 2020, the overall infection ratio of the US rises to 2.49\% 
while the case mortality decreases to 2.09\%, and the ratio of asymptomatic cases out of all infected cases reduces from the pre-summer 
35-40\% to around 20-25\%.
\end{abstract}


\section{Introduction}

\label{section:Intro}
The COVID-19 pandemic has caused major disturbance to human life. An important reason behind the widespread 
social anxiety is the huge uncertainty about the pandemic. One major uncertainty is how many or what percentage 
of people have been infected? There are published and frequently updated data on various statistics of the pandemic, 
at local, country or global level. However, due to various reasons, many cases were not included in those reported numbers. We 
propose a structured approach for the estimation of the number of unreported cases, where we distinguish cases that arrive late 
in the reported numbers and those who had mild or no symptoms and thus were not captured by any medical system at all. We use 
post-report data for the estimation of the former and population matching to the latter. We estimate that the reported number of infected cases 
in the US should be corrected by multiplying a factor of 220.54\% as of Apr 20, 2020. The infection ratio out 
of the US population is estimated to be 0.53\%, implying a case mortality rate at 2.85\% which is close to the 3.4\% suggested by the WHO
in March 2020.  
\\
\\
A flurry of work have appeared on the estimation of the number of infected or missing 
cases for COVID-19. One class of methods use the case fatality rate (CFR) as a proxy and then derive the number 
of infected cases from the death tolls \cite{GuptaShankar2020,JagodnikLachmann2020}. An accurate estimate 
of CFR is, however, challenging due to the use of different definitions in calculating the mortality counts in practice
and also the potential inflation in the reported case mortality \cite{JagodnikLachmann2020}. Indeed, estimation
through CFR may be misleading \cite{BottcherXiaChou2020}. Another class of methods are based on
epidemiology models such as the susceptible, infectious, recovered, and death (SIRD) model or its variants 
\cite{Richterich2020, TianWangZhang2020}. While the idea is fairly clean, these methods use heavy machinery such as 
differential equations and Markov chain Monte Carlo simulations which require nontrivial efforts for interpretation. 
Additionally, \cite{JagodnikLachmann2020} uses data from a benchmark country (South Korea) and then extrapolate 
by expected deaths and hospitalizations to the target country, while \cite{BaqueroPatras2020} resorts to
crowd sourcing, as an approximate way of random sampling, to estimate the ratio between the reported and the 
actual number of infected cases. These methods require the benchmark and the target country be 
similar in terms of cases and CFR, or a systematic way to control the sampling bias and the quality of crowd sourcing.   
\begin{figure}[h]
\centering
\begin{center}
\hspace{0cm}
\includegraphics[scale=0.36,clip]{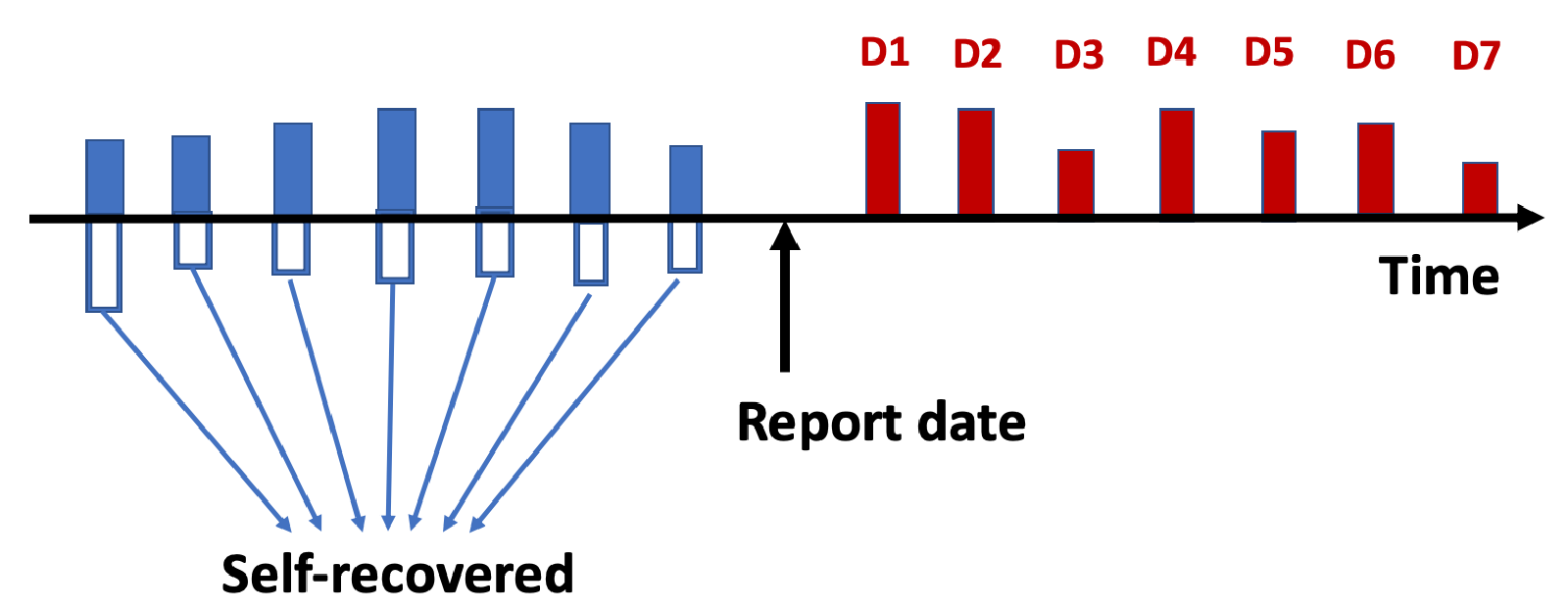}
\end{center}
\caption{\it Time train of the progression of infection with COVID-19. Blue or non-filled bars indicate daily new infected cases of type I or II, 
$D_1, ..., D_7$ are the number of reported cases during the 1\textsuperscript{th}, 2\textsuperscript{nd}, ..., and 7\textsuperscript{th} 
day past a given report date (indicated as ``Report date"). }  
\label{figure:timeTrain}
\end{figure}
\\
\\
In this work, we present a structured approach
for an approximate estimation of the number of infected cases at the US national and state level. 
Statistically, the estimation of the number of unreported cases is related to inference with missing data
\cite{LittleRubin2002} or censored data \cite{KaplanMeier1958}. However, certain characteristics of the coronavirus 
epidemiology allow us to take a different approach. Inspired by the diagnostic analysis 
of remote sensing studies \cite{rsDiagnosis2019} where the errors in the land use classification were decomposed 
according to their sources, we distinguish the missing counts in the reported numbers by two sources as illustrated 
by Figure~\ref{figure:timeTrain}.
One source is cases for which, at the time of report, the symptoms were not emerging yet; however, the affected 
individuals would eventually take a test with results reported. We call such cases the {\it type I cases}, and the waiting 
period before the onset of symptoms is termed as the {\it incubation (or dormant) period}. This is illustrated as the filled 
blue bars in Figure~\ref{figure:timeTrain}. At the time of a given report date (i.e., ``Report date" in Figure~\ref{figure:timeTrain}), 
all such cases are still in dormant status thus are missing in the reported number. A similar figure can be seen in the SIR 
compartment model considered by \cite{ZhouJi2020Arxiv}. Note the ``Report date" in Figure~\ref{figure:timeTrain} 
is just {\it a report date that is of interest}, and this is true henceforth unless otherwise specified. 
Usually there is a time gap between the onset of symptoms and the {\it time of report}---the infected individual may not immediately 
take the test and also there might be a delay in reporting (some reported statistics are based on the time of test though). 
In the lack of relevant data on such delays and for simplicity, we take an integer value slightly larger than the mean incubation 
periods and use that as the {\it time period to report} (i.e., the time period between infection and report, see 
discussion on the choice of the time window size in Section~\ref{section:delayedCounts}). 
\\
\\
The second source 
of unreported cases are those who were infected but are either not aware of it or with symptoms too light to bother, 
and later on recovered without any particular medical treatments. We call such cases the {\it type II cases}. The type 
II cases are never reflected in any reported numbers, thus leaving too little clue for estimation. But we cannot overlook 
such cases, since the number of such cases could be potentially large and would form an important source of infection. 
\\
\\
We use post-report data for the estimation of the number of type I cases, and population matching to that of 
type II cases. Due to the intuitive nature and the simplicity 
in implementation, our approach can be readily applied by the general public or the health department if they wish to obtain 
an approximate estimate of the actual number of infected cases in order to help understand the situation or to assist 
policy making and disease control.   
The remainder of this paper is organized as follows. In Section~\ref{section:approach}, we will describe our approach. This is
followed by a presentation of results in Section~\ref{section:exp}. Finally we conclude in Section~~\ref{section:conclusion}. 
\section{Methods}
\label{section:approach}
As stated in Section~\ref{section:Intro}, we take a structured approach by estimating the number of type I and 
II cases separately. These are described in Section~\ref{section:delayedCounts} and Section~\ref{section:popMatch}, 
respectively.
\subsection{Estimating the number of type I cases}
\label{section:delayedCounts} 
The estimation of the number of type I cases is facilitated by the following crucial observation. Though not included in the reported 
number while in the dormant period, such cases would eventually be exposed when the symptoms become so severe 
that the individuals have to seek medical treatments. By that time, those previously missed cases at the given 
report date (which was a few days ago) would be counted towards infected cases at some later report dates (though 
one would not know at which particular report date). Such numbers should be included at the given report date but 
surface only several days later; for this reason we call them {\it delayed counts}. If there is a way to estimate such 
delayed counts or their total, then one can estimate the number of type I cases for the given report date. 
\\
\\
It will be instructive to consider a simple {\it ideal} case where all infected cases have an incubation period of 7 days and 
there is no delay in test taking and reporting. In this ideal case, the numbers $D_1, D_2, ..., D_6$ in Figure~\ref{figure:timeTrain} 
are exactly the number of cases who were at their 6\textsuperscript{th},  5\textsuperscript{th}, ..., 1\textsuperscript{st} day 
of infection (i.e., the time between infection and the given report date), respectively, when counted at the given report date 
(i.e., ``Report date" in Figure~\ref{figure:timeTrain}). As the incubation period for all cases is 7 days, these are all 
the type I cases missed at the given report date but reflected perfectly later in the number of daily reported 
cases during the 6 days post-report {\it time window} (the window size is 1 day less than the incubation 
period). So the total number of type I cases at the given report date can be calculated by their sum, $\hat{D}_{type1}=\sum_{i=1}^6 D_i$.
\\
\\
The reality is, however, complicated. First, the incubation period (also the time period to report) varies for individual cases. 
Also, during the post-report time window, {\it newly infected} cases may arise and be reported due to their short incubation 
periods. Thus the number of daily reported cases at any particular day within this time window might be mixed, in the sense 
that it would include cases that are infected both before (but were during their incubation period) and after the given report 
date. The former case will not pose a problem as 
anyway such cases would be counted towards $\hat{D}_{type1}$ though cases infected on the same day may now 
contribute to different $D_i$'s. The latter case is undesirable but could be corrected, to a certain extent, by the truncating 
effect when we only sum up the daily counts in the post-report time window up to $T$ days. That is, those cases with 
a dormant period extending more than $T$ days post-report will be truncated and not included in $\hat{D}_{type1}$, with 
the total count of such truncated cases being `cancelled out' by the newly infected cases within the time window 
of length $T$. This leads to an {\it estimate} for the number of type I cases as
\begin{equation}
\hat{D}_{type1}=\sum_1^T D_i,
\label{eq:estimateTypeI}
\end{equation}
where $D_i$ are now the number of cases reported at the i\textsuperscript{th} day after the given report date. By intuition 
from the ideal case and also accounting for the potential delay in testing and reporting, we can let $T$ take a value 
around or slightly larger than the mean incubation period.
\\
\\
If we can keep track of the estimate $\hat{D}_{type1}$ through time, then we can get a time series which, upon 
smoothing, could be used to estimate the current count of type I cases. For such an estimation to be feasible, 
we have two assumptions. One is that the daily reported counts near the given report date would not change 
too abruptly. Thus, our approach may not work well when the infection trend rises very rapidly (e.g., during the 
initial outbreak of the pandemic). During such 
a period, the safest strategy might be to strictly enforce social distancing. 
The other is knowledge of the incubation period. A number of studies have been carried out on the estimation 
of incubation periods, for example, \cite{LauerGrantz2020, LintonKobayashi2020} report a median 
of 4--7 days, \cite{McAloonCollins2020} gives a mean of 5.8 days, \cite{BackerKlinkenbergWallinga2020} 
reports a mean of 6.4 days and 6.8 days by fitting the Weibull and lognormal distribution, respectively.
While further studies are required, we take $T=7$ in our estimation (which we believe also partially 
accounts for the time gap between the onset of symptoms and the time of report).
Additionally, it should be cautious that our estimation is valid assuming that the test of coronavirus is sufficiently 
carried out for the population of interest; insufficient test would render an underestimate. Here by sufficient testing 
we mean whoever or the vast majority of those with symptoms of COVID-19 would take the test. While this was 
hardly true during the initial outbreak of the pandemic due to resources constraints or public awareness of the 
pandemic, we believe gradually, at least for the US, it becomes reasonably true.    
\begin{figure}[htbp]
\centering
\begin{center}
\hspace{0cm}
\includegraphics[scale=0.5,clip,angle=-90]{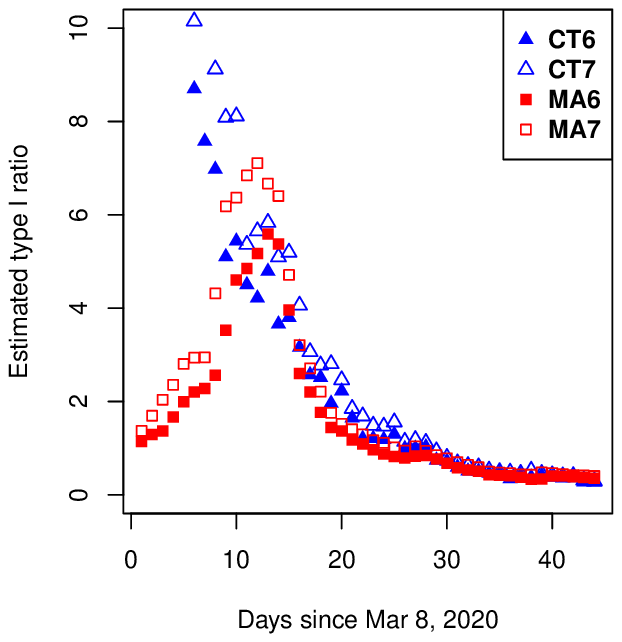}
\includegraphics[scale=0.5,clip,angle=-90]{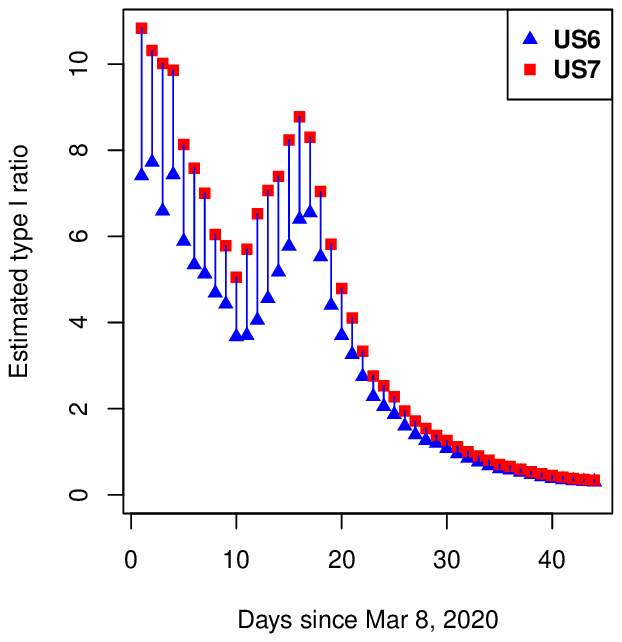}
\end{center}
\abovecaptionskip=-5pt
\caption{\it Ratio of unreported cases of Type I w.r.t. reported cases for CT, MA and the US, where `6' and `7' in the legend are the 
values of $T$ used, respectively. }  
\label{figure:unSeen2state}
\end{figure}
\\
\\
In Figure~\ref{figure:unSeen2state}, we plot the ratio of estimated type I cases w.r.t. the reported number of cases for 
Connecticut (CT) and Massachusetts (MA) since Mar 8, 2020. These two states were chosen as they are similar in many 
aspects, such as geography, demography, and population socioeconomics etc which are relevant in the spreading of COVID-19, 
so we expect their ratio of type I cases out of reported cases would be similar. In Figure~\ref{figure:unSeen2state}, 
there is an initial difference in ratios of type I cases in these two states, which we attribute to the late response and
the small number of cases tested in CT, and also possibly some random events such as the well-known {\it Biogen} 
superspreader event in MA in late February. Later, these two states exhibit strikingly similar trend, which is quite 
expected. We also explore the effect of using different values of $T$ where 6 and 7 are used. Again, initially the resulting 
estimation are quite different, which indicates the rapid spread of coronavirus and the rapid rise of infected cases. 
Gradually, the difference in the resulting estimations diminishes, which implies that the choice of $T=7$ leads to a 
fairly stable estimation after the initial quick growing stage. Similar observation can be made for the estimation of 
type I cases in US. This is shown in the right panel of Figure~\ref{figure:unSeen2state}. 
In the appendix, we give a statistical 
justification on why our estimate, $\hat{D}_{type1}$, would be a reasonable one. In particular, we derive an upper bound 
for the error of this estimate relative to the reported number of infected cases and show that it would be small under mild 
assumptions about the distribution of the incubation periods and the number of daily new cases. 
\subsection{Estimating the number of type II cases}
\label{section:popMatch}
The estimation of the number of type II cases is challenging, as there is barely anything observable on the asymptomatic cases. 
Our main strategy is based on the {\it matching of population statistics}---using what we see well to infer what is missing 
or incomplete. When grouping reported infection cases, we notice a significant discrepancy in the count statistics by 
different age groups in the US. We expect that, while people in most age groups in the population have a similar chance 
of being infected, infected individuals of age 65+ are very likely to be discovered timely. This is because people at age
65+ typically have a relatively weaker immune system along with possible pre-existing medical conditions, and a slight 
symptom would prompt them to seek medical treatments. Indeed the case mortality rate increases exponentially with age,
and could be well above 20\% for seniors over 70 
\cite{YanChenYangArXiv2021}. Thus reported counts about age groups 65+ would be more 
accurate and can serve as a {\it reference} to correct counts for other age groups. On the contrary, cases for the 20-64 age 
group are easy to be overlooked or unnoticed, thus their reported counts require a correction (termed as {\it age correction}). 
\begin{table}[h]
\begin{center}
\begin{tabular}{r|rrrrrrr}
\hline
 Age groups                      		& \bf{0-19}     	& \bf{20-44}   	&  \bf{45-54} &\bf{55-64} &\bf{65-74} &\bf{75-84} &\bf{85+}\\
\hline \hline
US population  		&25.06 	&33.27  &12.73  &12.92   &9.32      &4.70   	&2.00\\
Reported cases    & 5.00	&29.00  &18.00  &18.00  	&17.00	&9.00	&6.00\\
\hline				
Corrected cases		&46.47	&61.70	&23.61  &23.96 &17.00	&9.00	&6.00\\
\hline
\end{tabular}
\end{center}
\caption{\it Percentages by age groups in the US population (2020) and in the reported infection cases. Note numbers
in the bottom row are not normalized to sum up to 1. } \label{table:ageGroups}
\end{table}
\\
The age group of 85+ is more vulnerable to infections, as they typically live in the senior centers or extended-care nursing facilities 
which, as a matter of fact, have a very high risk of infection. The case statistics for this age group would be very thorough, 
but many in this age group get infected simply because they share a very confined living space with {\it many} other equally 
vulnerable seniors, and the infection of any one in a senior center may quickly spread to the rest ({\it to certain 
extent, one may think of this as a big indoor party during the pandemic}). 
So statistics in this age group would not be a reliable reference for population match, since people in other age groups have a 
very different mobility pattern (the infants interact with the world through their parents thus have a chance of infection 
not so different from the general population).  
\\
\\
As a result of equal age susceptibility for people with an age in the range 0-84, the number of infected cases of different age 
groups would be proportional to their respective percentage in the population.
Let $r_{pop}$ and $r_{case}$ be the proportions of the reference group in 
the population and in the reported cases, and $x_{pop}$ and $x_{corrected}$ be the the respective proportions for the target
group, respectively. Then 
\begin{equation*}
\frac{r_{case}}{r_{pop}} = \frac{x_{corrected}}{x_{pop}},
\end{equation*}
and the corrected percentage in the infected cases for the target group can be calculated accordingly.
As we argue before, the case statistics for age groups 65-74 and 75-84 are reliable so they are used as the reference group. 
A simple calculation reveals that age groups, 65-74 and 75-84, according to Table~\ref{table:ageGroups}, 
have a similar ratio of {\it cases percentage: population percentage}, i.e., $9.00:4.70 \approx 17.00:9.32$. Thus, we can pool 
counts from these two groups and obtain 
\begin{equation*}
r_{case} : r_{pop} = (9.00+17.00) / (4.70+9.32)=1.8544.
\end{equation*} 
This yields the corrected ratio as the bottom row of Table~\ref{table:ageGroups}. Adding up numbers in the bottom row gives 
a total of 187.94\%, implying that we should expand the reported counts by 87.94\% in order for the reported case counts to 
match the population statistics across age groups. This gives the ratio of type II cases over the reported cases. 
\\
\\
An interesting question is, will estimated counts of type II overlap with that 
for type I cases? We claim that this will not, at least not significantly, so the addition of estimated counts for type I 
and II cases is valid. The reason is that, type I cases still contribute to the reported numbers, at a delayed time though. These 
delayed cases can be thought of as a sample from the reported cases (assuming that the reported cases have a stable proportion 
when breakdown by age groups). The inclusion of type I cases will not change the age-breakdown proportions. Thus, after the 
inclusion of type I cases, we still have the same age-breakdown proportions and thus require an age correction. 
\subsection{Discussions on implications and assumptions}
In this section, we will briefly discuss the implications of type I and II ratios and the main assumptions made in our estimation.
\\
\\
The value of type I and II ratios can have important implications to understand the trend of the pandemic or for policy making. 
If there is no major 
re-surge of cases, then the type I ratio will slowly decrease with time as a result of the increasing total number of reported cases.
Thus as the pandemic continues, type II cases will gradually become the main source of unreported cases.
One interesting pattern about the dynamics of type I ratios with time is that a large value or an increase in type I ratio 
would indicate a quickly growing trend or a re-surge of the pandemic; this can be seen from Figure~\ref{figure:unSeen2state} by
the large value of type I ratios in the beginning. It can serve as a strong signal for policy makers or the health department to take 
immediate actions and for the public to be cautious. 
\\
\\
Type II cases are particularly harmful as they are asymptomatic, 
so it is always highly desirable to reduce the type II ratio. An effective way for this is to increase the coverage of the COVID-19 tests, 
and to enroll as many individuals as possible (subject to testing capacity) to take the test. Since the summer, many schools or colleges 
started introducing the asymptomatic test, and we think that has been very effective in helping reduce the type II ratio as these two 
groups contribute quite substantially to the asymptomatic cases. Also, COVID-19 tests related to travel and the re-opening of many 
states have helped detect many asymptomatic cases, due to the mandate testing of engaged individuals. Additionally, as more and 
more individuals are infected, all their close contacts are required to take the test although many are infected but not developing any 
symptoms. Such cases could be huge due to the exponential social network effect \cite{EasleyKleinberg2010}. Studies, including 
our analysis on more recent data (see appendix), show that the ratio of asymptomatic cases out of all infected ones has reduced 
from the pre-summer 35-40\% to around 20-25\% towards the end of the summer.
\\
\\
In estimating the type I ratio, we assume a stable distribution of the daily new infected cases. This may not be realistic during the 
initial outbreak of the pandemic or later sudden spikes of new cases, and our estimate will potentially overestimate during 
days around that period. On the other hand, during such period, likely many cases may not be captured by the reported counts. 
This leads to some cancellation effect to the overestimate, but to what extent is a complicated issue that requires additional
information. According to our sensitivity analysis (c.f. Section~\ref{section:sensitivity}), our estimate is fairly robust and incurs a small
error when there is a sudden increase of daily new cases by up to 10 times.
\\
\\
One major assumption in estimating the type II ratio is the equal susceptibility across age groups. This is a common assumption 
made by SIR models for which ages are not explicitly modeled, and we assume this for simplicity. People in age group 0-84 are 
either exposed to the infection directly, or indirectly through their family members (the chance for household infections is very high), 
thus it is reasonable to assume that people in this age group have roughly the same chance of infection during the pandemic. Indeed 
some studies \cite{BiWuMei2020} show that the susceptibility is not sensitive to ages. We note that some work in the literature 
\cite{ZhangLitvinovaLiang2020, PremLiuRussell2020} explicitly models the age-specific susceptibility. However, they either use 
the number of contacts for people in dense cities such as Wuhan and Shanghai in China during a single day as a proxy for 
age-specific susceptibility \cite{ZhangLitvinovaLiang2020}, or use simulations by locations such as working places, schools or communities etc 
to estimate age-specific contacts \cite{PremLiuRussell2020}. These studies omit the degrees of contacts and the levels of protections, 
and explore susceptibility in a single day (which departures from our goal, whether a person may eventually be 
infected during the pandemic). Moreover, their findings are not applicable to the US as the contact patterns among individuals would 
be very different, due to huge differences in life and working styles, population density or social distancing policy etc. We believe the 
difference among different age groups would be small for a sizable population, and the equal susceptibility assumption will facilitate a 
simple estimation. As many factors potentially contribute to the infection and spread of COVID-19, it is challenging to estimate the age 
effect but it is a worthwhile problem for further study, and clarifying it may help our estimation and many SIR-based models. 
\section{Results}
\label{section:exp}
\begin{figure}[htbp]
\centering
\begin{center}
\hspace{0cm}
\includegraphics[scale=0.6,clip,angle=-90]{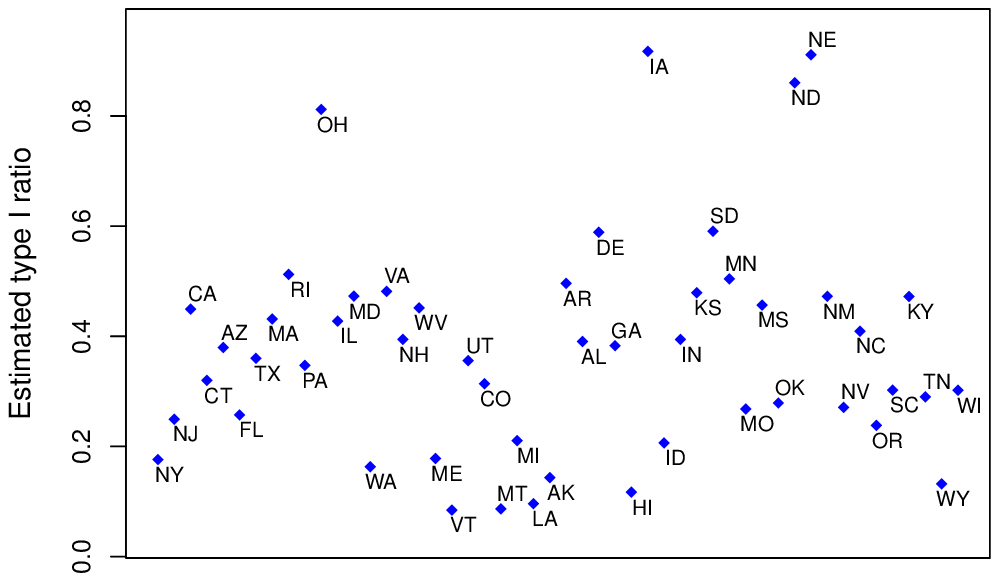}
\end{center}
\abovecaptionskip=-5pt
\caption{\it Ratio of type I cases w.r.t. reported cases for each of the 50 states in US on Apr 20, 2020. }  
\label{figure:unSeen50Type1}
\end{figure}

\begin{figure}[htbp]
\centering
\begin{center}
\hspace{0cm}
\includegraphics[scale=0.6,clip,angle=-90]{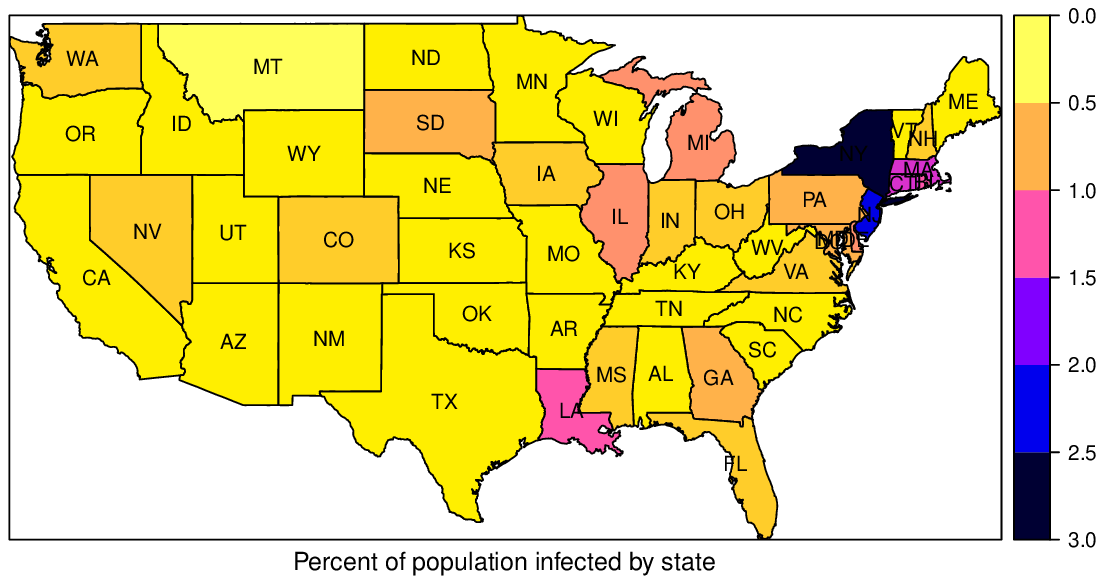}
\end{center}
\abovecaptionskip=-5pt
\caption{\it Heatmap of infection ratio for individual states as of Apr 20, 2020. }  
\label{figure:heatmapInf50}
\end{figure}

\begin{figure}[htbp]
\centering
\begin{center}
\hspace{0cm}
\includegraphics[scale=0.5,clip,angle=-90]{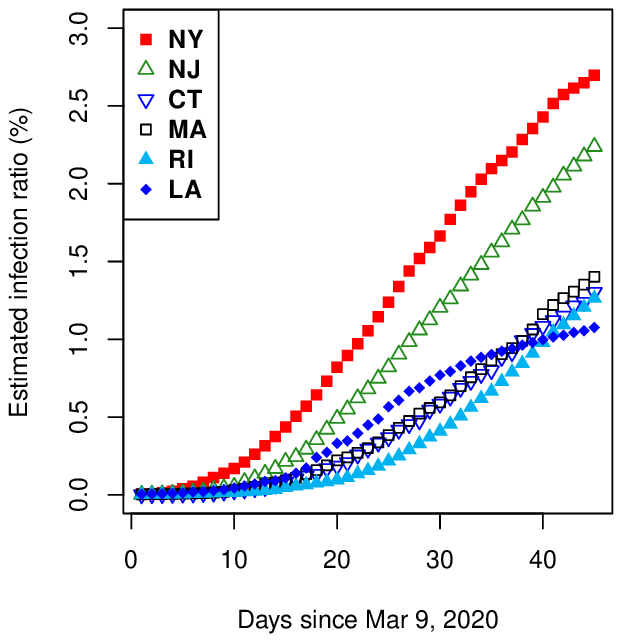}
\includegraphics[scale=0.5,clip,angle=-90]{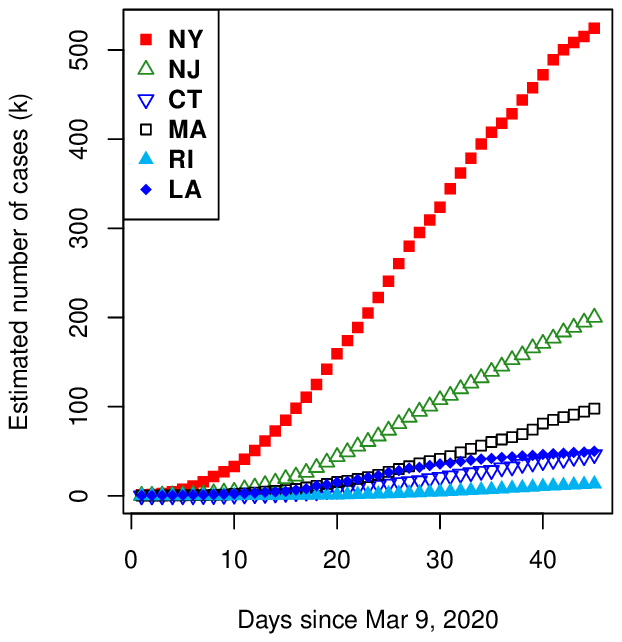}
\end{center}
\caption{\it Trend of estimated infection ratio and infected cases for NY, NJ, CT, MA, RI, and LA since Mar 9, 2020. }  
\label{figure:trendInf6}
\end{figure}

We apply our approach to each of the 50 states and the US. The data are available from Wikipedia \cite{WikiCovid19US}. 
Due to the large variation of the population size at different states, we calculate the {\it ratio of missing cases out of the 
number of reported cases} for individual states. The estimated ratios for type I cases for individual states, as of Apr 20, 2020, 
are shown in Figure~\ref{figure:unSeen50Type1}. 
The type I ratio for OH, IA, ND and NE are substantially higher than others. This because the infected cases were still quickly rising 
for OH, IA and ND around the time the type I ratio was calculated. For NE, the accumulated number of reported cases is 
small so the addition of new daily reported cases may easily lead to a high type I ratio. 
It should be noted that the ratio of type I cases out of the reported number will decrease as the number of daily reported 
cases begins stabilizing. This can be seen from Figure~\ref{figure:unSeen2state}.
Due to the lack of reported case data by age groups for some individual states, we use the overall estimate for the US, 
which is 87.94\% according to discussions in Section~\ref{section:popMatch}, for the ratio of type II cases for all 
the states.
\\
\\
The overall ratio of type I cases for the US is estimated to be 32.60\% by aggregating estimated type I cases from 
individual states. Combining with the type II case ratio at 87.94\%, this gives an estimated ratio of missing cases versus 
the reported number at 120.54\% for the US. In other words, the reported number of cases for the US should multiply 
by a factor of 220.54\% to reflect the true number of infected cases.
This is close to the estimate given in \cite{EwaldMyrskyla2020} which is twice as large as the number of reported 
cases for the US as of Apr 17, 2020. Our result implies that the proportion of cases 
never captured by any medical systems out of all the infected cases is about 87.94/220.54=39.87\% as of 
Apr 20, 2020. This is not far from estimations given by \cite{TianWangZhang2020} which predicates that the ratio 
of unidentified cases in NY, NJ, and CA by July 11, 2020 in the range between 18.49\% and 33.20\%. Our estimate
is also consistent to a meta-analysis of over 10 studies \cite{Byambasuren2020} which reports an asymptomatic ratio 
in the range of 15\%-40\%.
\\
\\
With the unreported numbers estimated, we can estimate the {\it infection ratio}, defined as {\it the ratio
of the number of infected cases out of the population}. The overall infection ratio of the US is estimated to be 0.53\%, 
or 1.75 million infected cases, as of Apr 20, 2020. If we use the associated death toll at about 50k, then 
the case mortality rate is calculated as 2.85\%, which is close to the WHO suggested estimation of 3.4\% \cite{WHOCOVIDMortality} 
in March, 2020. 
The infection ratios for individual states are visualized as heatmap in Figure~\ref{figure:heatmapInf50}. Heavily hit states 
are NY, NJ, CT, RI, MA, and LA with infection ratio estimated at 2.61\%, 2.11\%, 1.22\%, 1.15\%, 1.31\% and 1.04\%, 
respectively, as of Apr 20, 2020. The trend of the infection ratio and cases over time for these states is shown in Figure~\ref{figure:trendInf6}. 
It can be seen that, except LA, the infection ratios for all other five states are still rapidly 
increasing. NJ shows a similar growing pattern as NY, while the three New England states, CT, RI and MA, are similar. 
\subsection{Sensitivity analysis}
\label{section:sensitivity}
Due to the lack of true infected case counts, we are not able to evaluate the estimation error of our approach empirically. 
Instead we carry out a sensitivity analysis by simulations, with the goal of evaluating the robustness of our approach under 
fluctuations of daily new cases. The simulations are conducted as follows. On each day over a span of about 100 days, we 
generate $N_i$ new infected cases with incubation periods following the lognormal or the Weibull distribution. In a given 
future date, all cases whose incubation period expires will be reported (note cases infected on different dates may be 
reported on the same day or infected on the same day may be reported on different future dates). On each day, we then 
compare the actually reported (or unreported) cases to the estimated cases of type I by our algorithm. 
\\
\\
The incubation periods are generated according to lognormal \cite{McAloonCollins2020} or Weibull \cite{BackerKlinkenbergWallinga2020}. 
We follow the references in their choices of parameters. The lognormal distribution 
has 1.63 and 0.5 as its mean and standard deviation on the log scale, or 5.80 and 3.08 days for mean and standard 
deviation of incubation periods. The shape and scale parameters for the Weibull distribution are 
3.04 and 7.16, or 6.4 and 2.3 days for the mean and standard deviation, respectively. 
\\
\\
To simulate the fluctuation of daily infected cases, we consider four scenarios. One is that the number of 
daily infected cases is kept at a constant, say, 1000. For the other three cases, we let the number of daily infected cases 
be sampling from intervals, [600, 1200], [400, 1600], and [200, 2000], respectively, such that the number of new 
cases can potentially be 2, 4 or 10 times larger, on a certain day, than that of a previous day. Some correlation structures 
can be imposed on the number of cases during consecutive days, but we opt for simplicity and choose not to pursue it
as we aim at simulating different levels of fluctuations of daily new cases.
The window size would be 4.8 and 5.4 days for lognormal and Weibull, respectively, in ideal case. But we have to pick an integer
value, and also we may need to compensate the potential delays with testing and reporting in practice. So we use window size 5 
or 6 for lognormal, and 6 or 7 for Weibull in our simulation. Note that the choice of a large window size may lead to an overestimate 
of the type I ratio. 
\begin{figure}[htbp]
\centering
\begin{center}
\hspace{0cm}
\includegraphics[scale=0.39,clip,angle=-90]{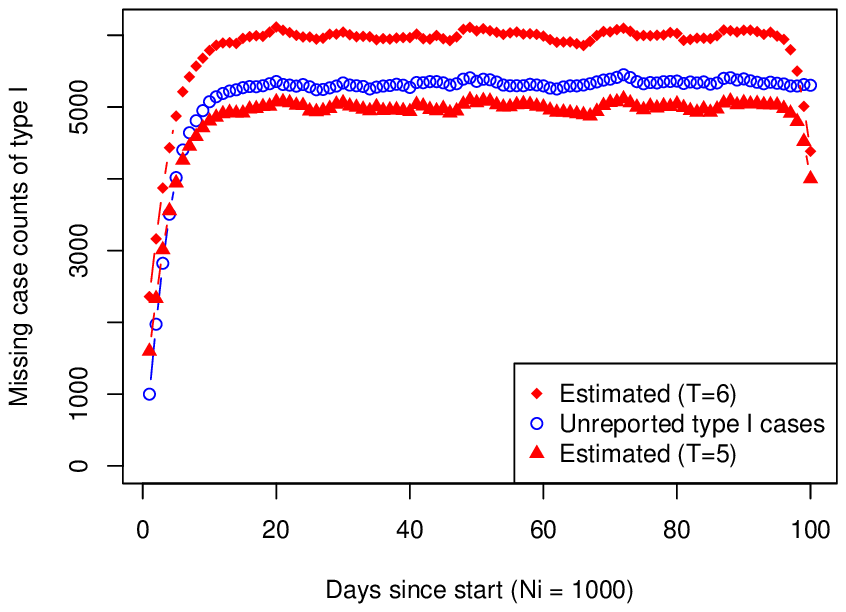}
\includegraphics[scale=0.39,clip,angle=-90]{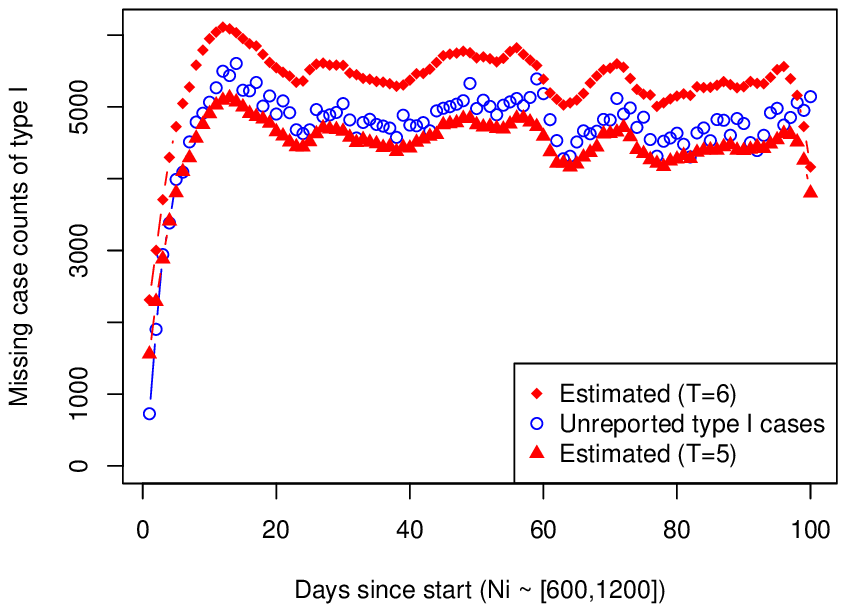}
\includegraphics[scale=0.39,clip,angle=-90]{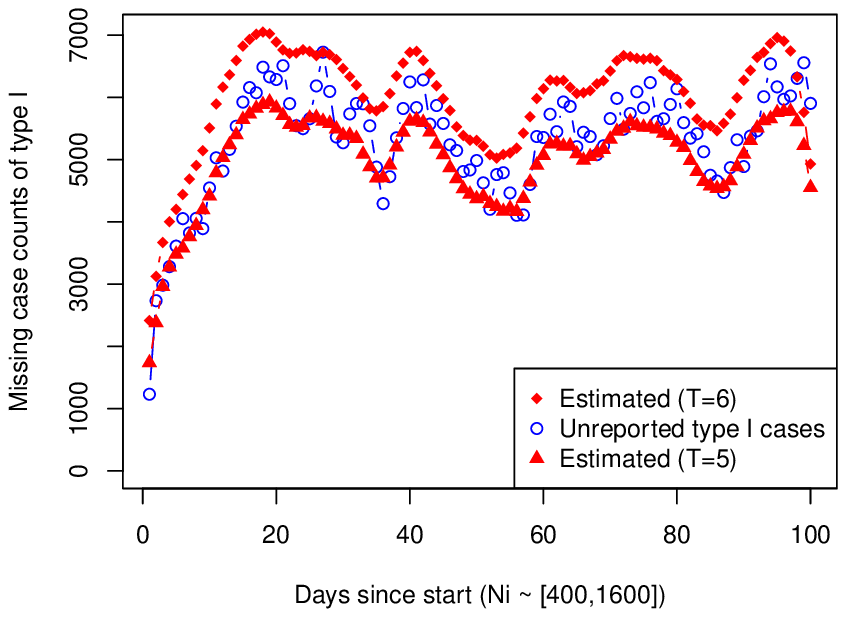}
\includegraphics[scale=0.39,clip,angle=-90]{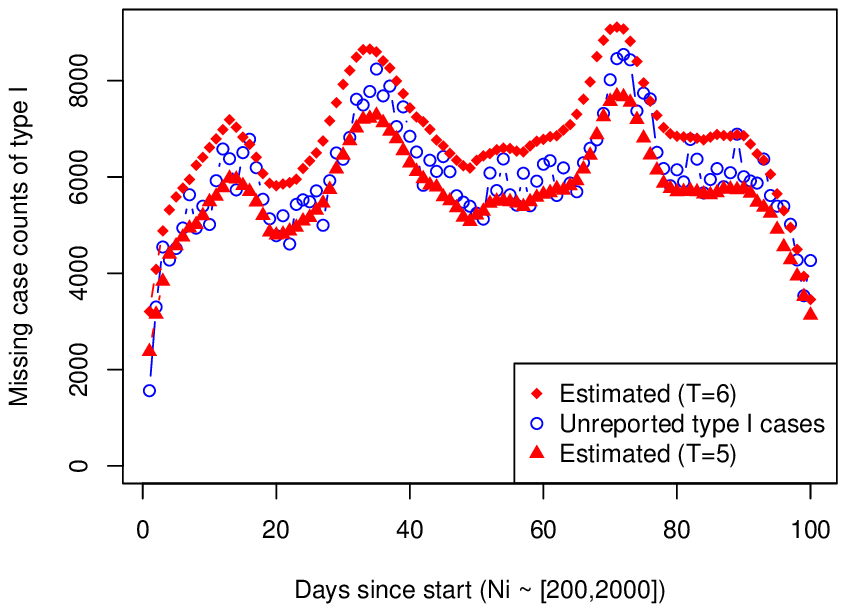}
\end{center}
\abovecaptionskip=-5pt
\caption{\it True and estimated number of type I cases with lognormal incubation periods. The window 
size $T$ used in the estimation are 5 and 6, respectively. The daily infected cases $N_i$ are generated 
under 4 scenarios, constant 1000, uniformly from range [600,1200], [400,1600], and [200,2000], respectively. }  
\label{figure:sensitivityL}
\end{figure}
\begin{figure}[h]
\centering
\begin{center}
\hspace{0cm}
\includegraphics[scale=0.39,clip,angle=-90]{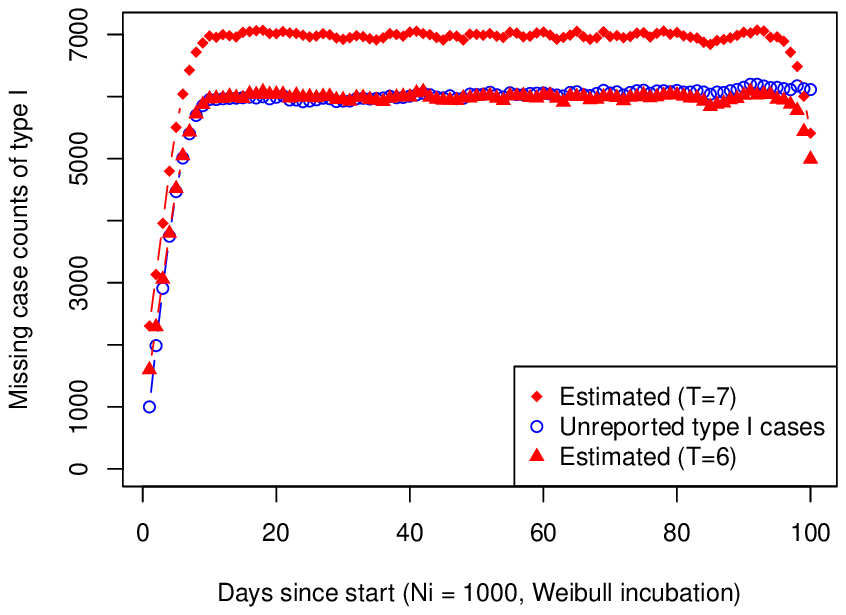}
\includegraphics[scale=0.39,clip,angle=-90]{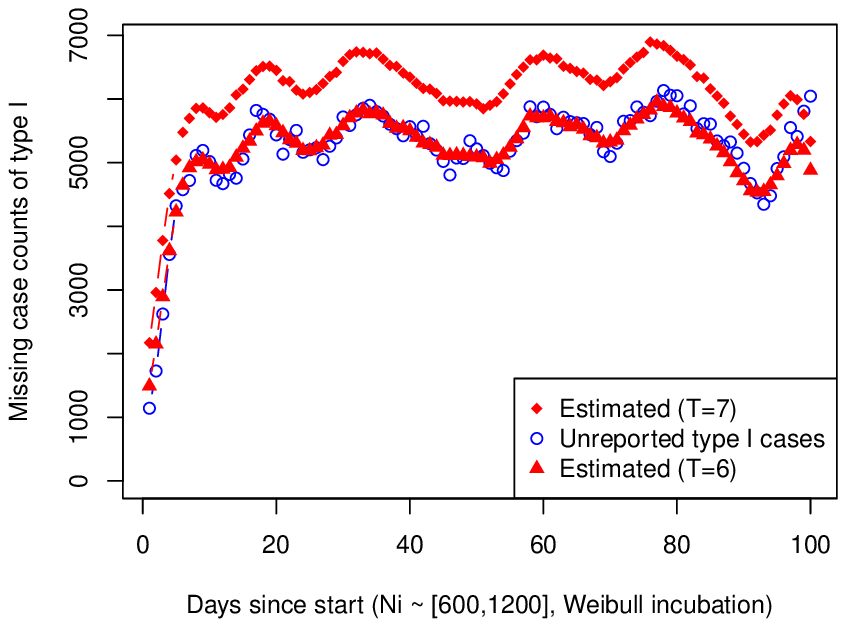}
\includegraphics[scale=0.39,clip,angle=-90]{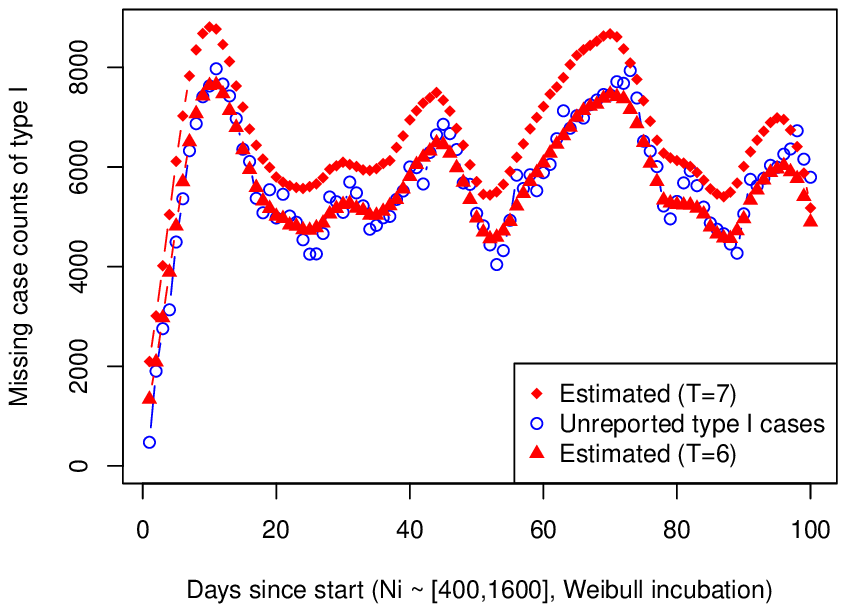}
\includegraphics[scale=0.39,clip,angle=-90]{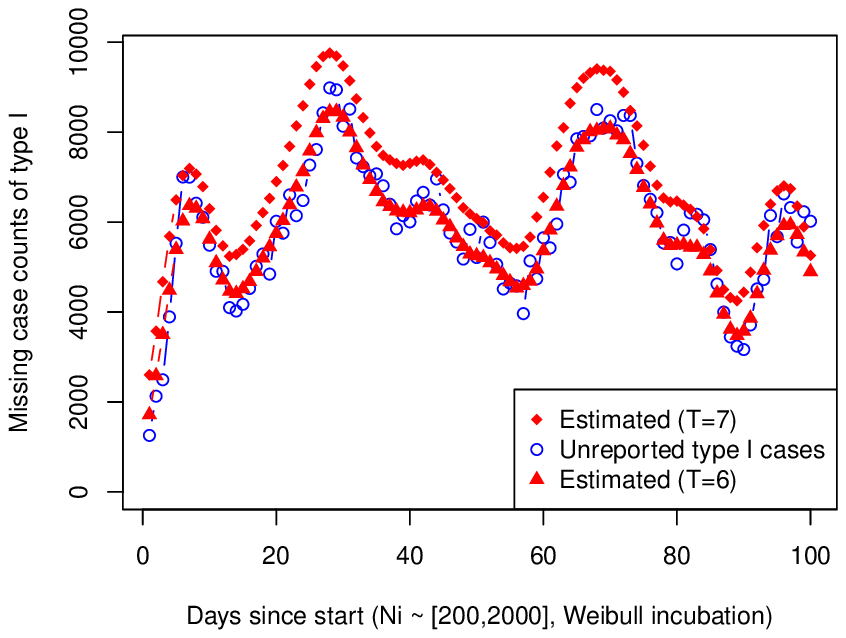}
\end{center}
\abovecaptionskip=-5pt
\caption{\it True and estimated number of type I cases and estimated with Weibull incubation periods. 
The window size $T$ used in the estimation are 6 and 7, respectively. The daily infected 
cases $N_i$ are generated under 4 scenarios, constant 1000, uniformly from range [600,1200], [400,1600], and [200,2000], respectively. }  
\label{figure:sensitivityW}
\end{figure}
\\
\\
Under all four scenarios, we plot the actually number of unreported cases vs the estimated ones. Figure~\ref{figure:sensitivityL} 
and Figure~\ref{figure:sensitivityW} are for the lognormal and Weibull distribution, respectively. We also calculate the percentage 
of errors w.r.t. the number of reported cases and average over 100 days and 100 runs. 
This is shown in Table~\ref{table:sensitivity}. 
\begin{table}[htbp]
\begin{center}
\footnotesize
\begin{tabular}{r||c|c|c|c}
\hline
                   	& $N_i=1000$     	& $N_i \sim [600, 1200]$  	&  $N_i \sim [400,1600]$  &$N_i \sim [200,2000]$\\
\hline\hline
Lognormal (T=5)  		&1.29\%		&1.34\%          &1.60\%	 &1.83\%\\
Lognormal (T=6)         &3.11\%             &3.14\%	       &3.17\%	 &3.22\%\\
\hline
Weibull (T=6)  		&0.38\%		&0.81\%          &1.34\%	 &1.75\%\\
Weibull (T=7)         &5.02\%             &5.08\%	       &5.15\%	 &5.27\%\\
\hline
\end{tabular}
\end{center}
\caption{\it Mean percentage of errors w.r.t. the number of reported cases over a span of 100 days. }
\label{table:sensitivity}
\end{table}
\normalsize
In all cases, the errors are small even when the number of cases fluctuates wildly from time to time. 
For a smaller window size (i.e., 5 for lognormal and 6 for Weibull), the errors range from 1.29\%--1.83\% 
and 0.38\%--1.75\% for lognormal and Weibull incubation, respectively. For a larger window size (i.e., 6 for 
lognormal and 7 for Weibull), the errors range from 3.11-3.22\% and 5.02--5.27\% for lognormal and Weibull, respectively. 
\section{Conclusions}
\label{section:conclusion}
We have proposed a structured approach for the estimation of the unreported number of infected cases. We distinguish two 
types of missing cases, those cases which were infected but are still during 
the dormant period at the time of report and those asymptomatic or light cases which later self-recover without any medical 
treatments. The number of these two types of cases are estimated by accumulating reported counts within a properly 
chosen post-report time window and by population matching. The reported number, as of Apr 20, 2020, of infected cases 
in US should be corrected by multiplying a factor of 220.54\%. The overall infection ratio out of the US population is estimated 
to be 0.53\%, implying a case mortality rate of 2.85\% as of Apr 20, 2020. The estimate given by our approach also agrees or 
is close to related estimates by several work in the literature. By Aug 31, 2020, the overall infection ratio in US rises to 
2.49\% while the case mortality rate decreases to 2.09\%, 
and the ratio of asymptomatic cases out of all infected cases reduces from the pre-summer 
35-40\% to around 20-25\%.
\\
\\
The intuitive nature of our approach makes it easy to understand and to implement, thus we expect it be readily adopted by the 
general public for understanding the situation or the government for policy making and disease control. Our estimation can 
potentially be used for risk assessment. The infant age group may worth further consideration as people in this group 
are much less risky than other age groups as they interact with the rest of the world through their parents, so the number 
of cases for this group may need to adjust accordingly to reflect the true risk. 


\section*{Acknowledgements}
We thank the Editor, Associate Editor, and referees for their constructive comments and suggestions.
We would also like to thank all those who contributed to the collection, curation and sharing of data related to COVID-19.
\par

\section*{Appendix}
\label{section:appendix}
\renewcommand\thesubsection{A\arabic{subsection}}
This appendix consists of two parts. In the first part,  we give a justification on our estimation algorithm for the number of type I cases.
In the second part, we include analysis on the US COVID-19 data as of August 31, 2020. 
\subsection{Error analysis in the type I estimator}
In this section, we show that 
the error between our estimate, $\hat{D}_{type1}$, of the number of type I cases and its actual value $D_{type1}$ is 
small in expectation under mild assumptions about the distribution of the incubation periods.
\\
\\ 
Denote by random variable $X$ the length of the incubation period, and for simplicity we further assume that $X\geq 0$ 
takes integer values. Let $N_{-1}, N_{-2}, ...$ denote the number of cases that were infected one day, two days and so 
on {\it before} the report date (for which we use $N_0$), while $N_i$'s indicate those after the report date. Here we limit 
to type I cases as we can conveniently assume that type II cases have an infinite incubation period. Then the expected 
number of cases that are discovered during the time window of $T$ days following the given report date is calculated 
as
\begin{equation}
D_a=\sum_{i=0}^{\infty} \mathbb{E}N_{-i} \cdot P(i+1 \leq X < i+T).
\label{eq:expectedCases}
\end{equation}
For simplicity, we would assume that all the $\mathbb{E}N_{-i}$'s take the same value $N$. This is a very mild assumption
as we expect that the distribution of the number of newly infected cases per day do not drift much when the pandemic 
reaches a stable stage (those at the very far distant past would be small, but they carry a very small fraction of the total 
number so could be ignored). Another simplifying assumption is the independence of $N_i$'s and the incubation period $X$. 
Also, we abuse the notation a bit by using $D_{.}$'s to also indicate the expected value of the associated random variable; 
the exact meaning will be determined by the context. Then Equation~\eqref{eq:expectedCases} can be rewritten as 
\begin{eqnarray*}
D_a &=& (T-1) \cdot N - N\sum_{i=1}^{T-1} (T-i) \cdot P(i-1 \leq X < i).
\end{eqnarray*}
Under the same assumption, the number of new cases generated during the post-report time window of length $T$ days is
\begin{eqnarray*}
D_{new} &=& \sum_{i=1}^{T} \mathbb{E}N_i \cdot P(X < T+1-i) \\
&=& N \sum_{i=1}^{T} (T+1-i) \cdot P( i-1 \leq X < i)\\
&=& N \sum_{i=1}^{T-1} (T-i) \cdot P( i-1 \leq X < i) + N \cdot P(X < T) 
\end{eqnarray*}
Thus the total number of reported cases during the post-report time window of length $T$ days is calculated as
\begin{equation*}
D_a+D_{new}= (T-1) \cdot N + N \cdot P(X \leq T) = TN - N \cdot P(X > T).
\label{eq:expectedCases3}
\end{equation*}
Assuming that random variable $X$ has a finite mean, say $\mu$, then we have
\begin{equation*}
P(X > T) \leq \mathbb{E}X/T = \mu/T,
\end{equation*}
implying that the estimated number of type I cases satisfies
\begin{equation}
TN \geq \hat{D}_{type1}=D_a+D_{new} \geq TN - N \cdot \mu/T.
\label{eq:InequalTypeI}
\end{equation}
The actual number of cases that have accumulated but not being discovered before the report date consists of
missing cases during the previous $T$ days and those even earlier cases, which has an expected value
\begin{equation}
D_{type1} = \sum_{i=0}^{\infty} \mathbb{E}N_{-i} \cdot P(X \geq i+1) = N \sum_{i=0}^{\infty} P(X \geq i) = N \cdot \mathbb{E}X = \mu N.
\label{equation:trueType1}
\end{equation}
\eqref{equation:trueType1} indicates that the mean number of type I cases equals the product of the mean daily infected 
cases of type I and the mean length of the incubation period, which is consistent with the ideal case discussed in Section~\ref{section:delayedCounts}. 
This, along with \eqref{eq:InequalTypeI}, suggests that the length of the post-report 
time window should be around the mean of the incubation periods. Let $T=(1+\epsilon)\mu$, then we have the following 
error bound for the estimated number of cases of type I 
\begin{equation*}
|\hat{D}_{type1} - D_{type1}| \leq \mu N \cdot \max(\epsilon, |\epsilon-1/T|).
\end{equation*}
It follows that the relative error of the estimate satisfies
\begin{equation*}
\frac{|\hat{D}_{type1} - D_{type1}|}{D_{type1}} \leq \max(\epsilon, |\epsilon-1/T|) 
= \max(\epsilon, |\epsilon-1/((1+\epsilon)\mu)|).
\end{equation*}
For a given $\mu$, one can pick $\epsilon$ to optimize the above bound. For example, when $\mu=7$, one can 
take $\epsilon=0.07$ to achieve a relative error bound of about 7\%. 
\par
\subsection{Analysis on more recent data}
\begin{figure}[htbp]
\centering
\begin{center}
\hspace{0cm}
\includegraphics[scale=0.60,clip,angle=-90]{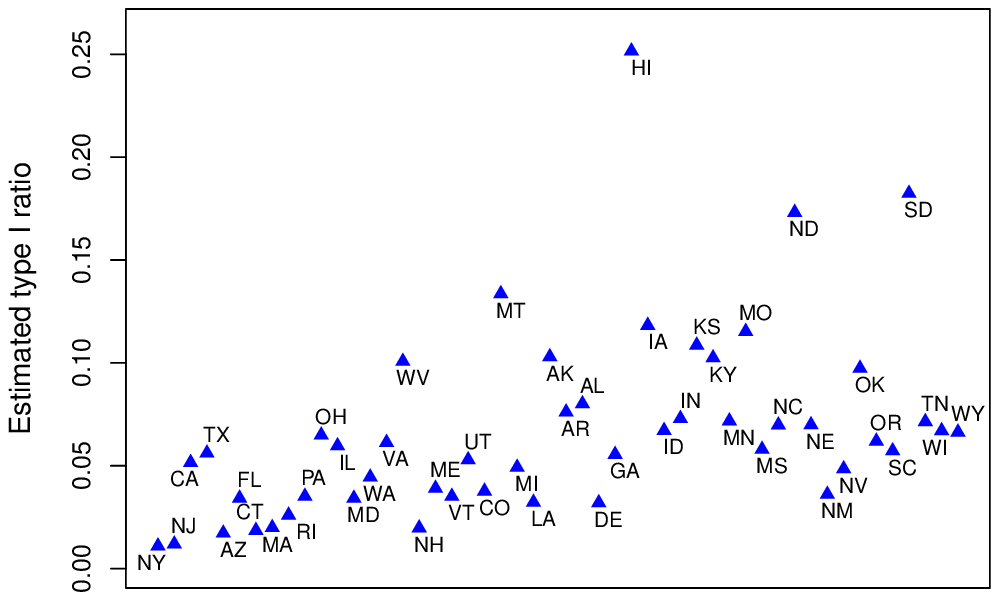}
\end{center}
\abovecaptionskip=-10pt
\caption{\it Ratio of type I cases w.r.t. reported cases for each of the 50 states in US as of Aug 31, 2020. }  
\label{figure:unSeen50Type1-0831}
\end{figure}

\begin{figure}[htbp]
\centering
\begin{center}
\hspace{0cm}
\includegraphics[scale=0.6,clip,angle=-90]{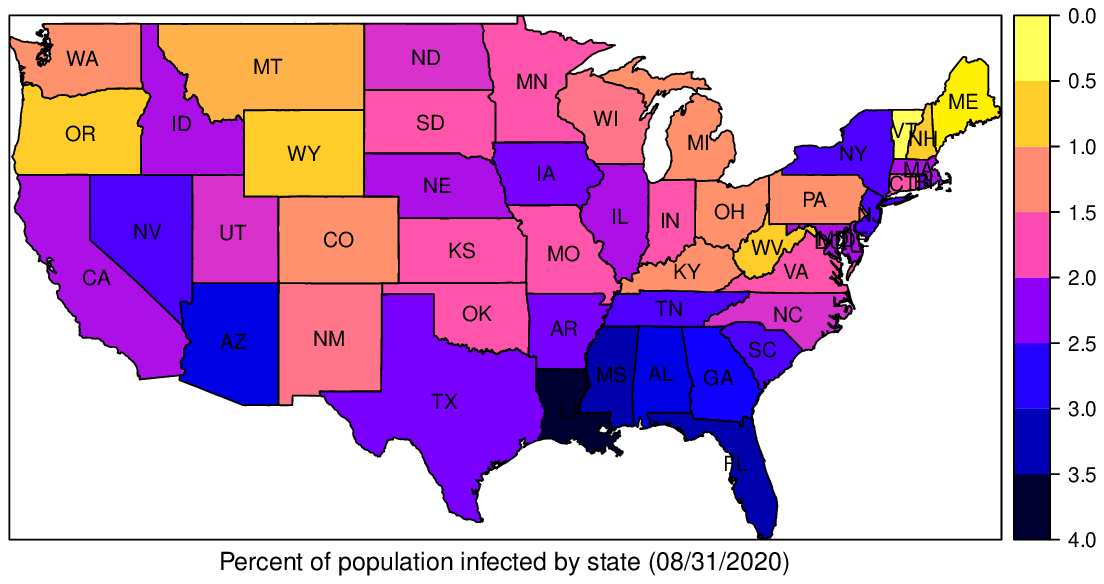}
\end{center}
\abovecaptionskip=-5pt
\caption{\it Heatmap of infection ratio for individual states as of Aug 31, 2020. }  
\label{figure:heatmapInf50-0831}
\end{figure}

We repeat our analysis on data as of August 31, 2020. 
The data consists in a surge of cases during the summer for many US states, with the reported case count for US increasing 
by several times from about 830k to 5.94 million since late April. For the new data, our analysis suggests that the 
number of reported cases for the US be adjusted by a factor of 138.56\%, with the ratio of type I and type II cases estimated 
to be 5.02\% and 33.54\%, respectively. Both ratios decrease significantly compared to estimates given by data up to 
Apr 20, 2020. The type I ratio decreases mainly as a result of a major increase in the total number of reported cases, 
from 830k to 5.94 million. The total number of infected cases in the US is estimated at 8,234,946 as of Aug 31, 2020, implying 
an overall infection ratio 
at 2.49\%. With a death toll of 171,957, this gives an estimated mortality ratio at 2.09\%, which lower than the 2.85\% in late Apr, 2020.
This is expected, and also observed in many other countries, either because we have now understand more about the virus and 
hence more informed medical treatments, or likely the virus has gradually become less lethal with time.
\\
\\
Similar as in Section~\ref{section:exp}, we show the estimated type I ratio for all 50 states 
as of Aug 31, 2020 (see Figure~\ref{figure:unSeen50Type1-0831}). Compared to Figure~\ref{figure:unSeen50Type1}, 
the type I ratios of all the states show a significant decrease, which indicates a major slow down of the spread of COVID-19. 
OH, IA, and NE no longer stand out while ND, along with HI, are still pretty high. 
\\
\begin{table}[h]
\begin{center}
\begin{tabular}{r|rrrrrrr}
\hline
 Age groups                      		& \bf{0-19}     	& \bf{20-44}   	&  \bf{45-54} &\bf{55-64} &\bf{65-74} &\bf{75-84} &\bf{85+}\\
\hline \hline
US population  		&25.06 	&33.27  &12.73  &12.92   &9.32      &4.70   	&2.00\\
Reported cases    & 10.84	&43.86  &15.48  &14.40  	&12.94	&5.77	&2.76\\
\hline				
Corrected cases		&33.44	&44.40	&16.99  &17.24 &12.94	&5.77	&2.76\\
\hline
\end{tabular}
\end{center}
\caption{\it Percentages by age groups in the US population (2020) and in the reported infection cases as of August 31, 2020. Note that numbers
in the bottom row are not normalized to sum up to 1. } \label{table:ageGroups2}
\end{table}
\\
The estimation of type II ratio is based on Table~\ref{table:ageGroups2}.
The decrease of type II ratio is likely because a lot more people have now taken the COVID-19 test, 
not necessarily due to the emerging of symptoms but because, as many states have reopened since the mid of May, people are 
required to take the 
test to resume working or to be engaged in summer travels or activities. Also as a lot more individuals are infected, all their 
close contacts are required to take the test even though many are infected but not showing any symptoms. Such cases could be 
huge due to the exponential social network effect \cite{EasleyKleinberg2010}. 
The percent of type II cases out of estimated total infected cases is estimated as 33.54\%/(1+38.56\%)=24.21\%. 
This is comparable to or consistent with a number of recent studies, for example, 
over 10 studies considered in a meta-analysis \cite{Byambasuren2020} report an asymptomatic rate in the range of 15\%-40\%.
With the introduction of asymptomatic tests in a large number of schools, colleges and the general public since the mid of summer, 
we expect that the type II ratio will further decrease over time and so will the overall ratio of unreported cases. 
\\
\\
Similarly, we also calculate the infection ratio and cumulative infected cases for the 6 states NY, NJ, CT, MA, RI 
and LA up to Aug 31, 2020. The similar trend for all other states pretty much continues except LA, which overtakes 
the rest and become the highest among all these 6 states; indeed it becomes the highest among all 50 states as 
well as of Aug 31, 2020, which is likely due to the aggressive re-opening since the mid of May. The top 6 states with 
highest estimated infection ratios are now 
LA (4.47\%), FL (3.97\%), MS (3.89\%), AZ (3.83\%), AL (3.60\%), and GA (3.52\%), while the previous (as of 
Apr 20, 2020) top 6 states have estimated infection ratio as follows: NY (3.14\%), NJ (3.02\%), CT (2.09\%), MA (2.58\%), RI (2.92\%) 
and LA (4.47\%). Similarly, we also include the heatmap of infection ratios of all 50 US states as of Aug 31, 2020 in Figure~\ref{figure:heatmapInf50-0831}.
\begin{figure}[htbp]
\centering
\begin{center}
\hspace{0cm}
\includegraphics[scale=0.4,clip,angle=-90]{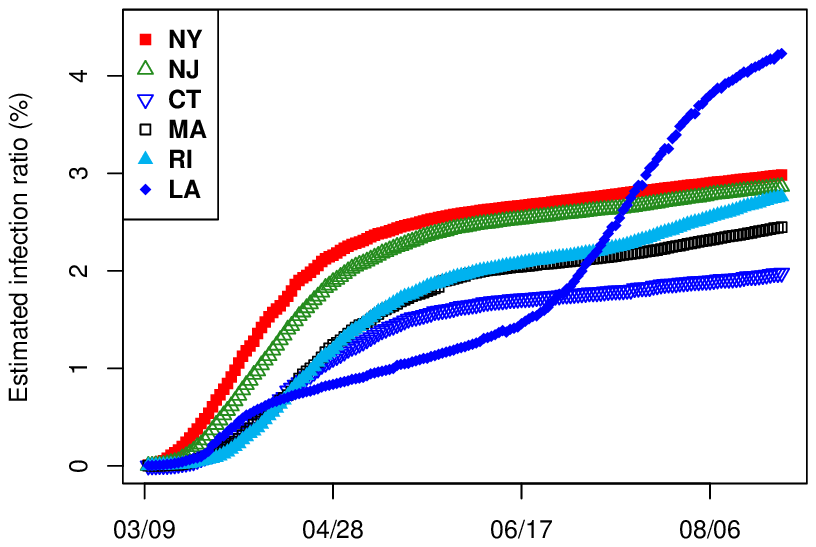}
\includegraphics[scale=0.4,clip,angle=-90]{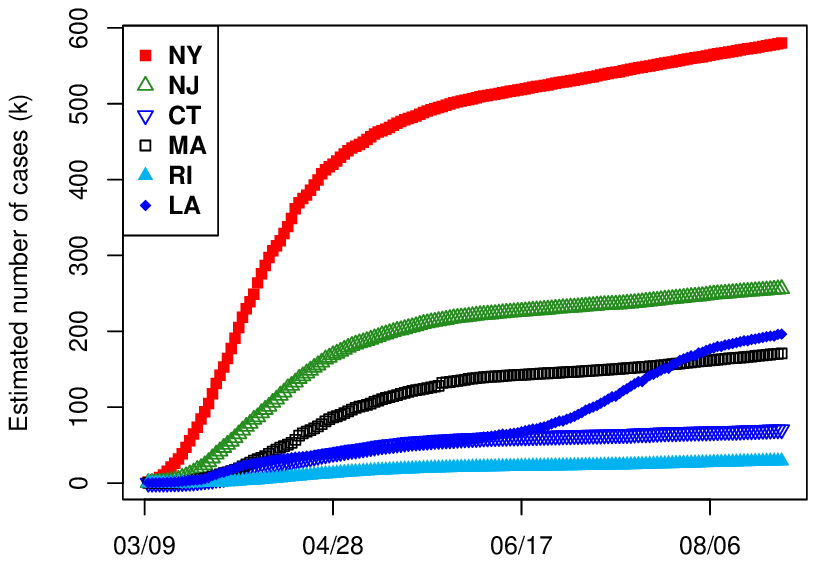}
\end{center}
\abovecaptionskip=-5pt
\caption{\it Trend of estimated infection ratio and infected cases for NY, NJ, CT, MA, RI, and LA since Mar 9, 2020. }  
\label{figure:trendInf6-0831}
\end{figure}
\par


\end{document}